\documentclass[iop]{emulateapj}
 \usepackage{apjfonts}
\usepackage{amsmath,bm}
\usepackage{multirow}
\usepackage{enumerate}
\usepackage{comment}
\usepackage{url}
\usepackage[colorlinks, citecolor=blue]{hyperref}

\begin{document}

\title{Catastrophe versus instability for the eruption of a toroidal
       solar magnetic flux rope}

\author{B. Kliem$^{1,2,3,4}$,
        J. Lin$^{1}$,
        T. G. Forbes$^{5}$,
        E. R. Priest$^{6}$,
    and T. T\"or\"ok$^{7}$}

\affil{$^1$Yunnan Observatories, Chinese Academy of Sciences,
                 Kunming 650011, China}
\affil{$^2$Institute of Physics and Astronomy, University of Potsdam,
                 Potsdam 14476, Germany}
\affil{$^3$Mullard Space Science Laboratory, University College London,
                 Holmbury St.\ Mary, Dorking, Surrey, RH5 6NT, UK}
\affil{$^4$College of Science, George Mason University, Fairfax, VA
                 22030, USA}
\affil{$^5$Institute for the Study of Earth, Oceans, and Space,
                 University of New Hampshire, Durham, New Hampshire 03824, USA}
\affil{$^6$School of Mathematics and Statistics, University of
                 St Andrews, North Haugh, St Andrews KY16 9SS, UK}
\affil{$^7$Predictive Science Inc., 9990 Mesa Rim Rd., Ste 170,
                 San Diego, CA 92121, USA}

\shorttitle{Catastrophe vs.\ instability}  
\shortauthors{Kliem et al.}

\email{bkliem@uni-potdam.de}

\journalinfo{Manuscript ms\_cat, revised 2014 April 17} 
\submitted{Received 2014 February 12; accepted 2014 April 22}

\begin{abstract}

\noindent  The onset of a solar eruption is formulated here as either
a magnetic catastrophe or as an instability. Both start with the same
equation of force balance governing the underlying equilibria. Using a
toroidal flux rope in an external bipolar or quadrupolar field as a
model for the current-carrying flux, we demonstrate the occurrence of
a fold catastrophe by loss of equilibrium for several representative
evolutionary sequences in the stable domain of parameter space. We
verify that this catastrophe and the torus instability occur at the
same point; they are thus equivalent descriptions for the onset
condition of solar eruptions.

\end{abstract}


\section{Introduction}
\label{s:introduction}

The force-free equilibrium of a coronal magnetic flux rope that
carries a net current requires the presence of an external poloidal
field perpendicular to the current \citep{Shafranov1966,
vanTend&Kuperus1978}. Magnetic flux associated with the current is
squeezed between the current and the photospheric boundary. This can
be described as an induced current in the boundary or, equivalently,
as an oppositely directed image current, implying an upward Lorentz
force on the coronal flux \citep{Kuperus&Raadu1974}. The force is
balanced by a Lorentz force from the external poloidal field.

As the photospheric flux distribution and the corresponding external
field gradually change, the configuration evolves quasi-statically
along a sequence of stable equilibria for most of the time. However,
it may encounter an end point of such a sequence, where continuing
photospheric changes trigger a dynamic evolution. The transition of an
equilibrium flux rope to a state of non-equilibrium has become a
standard model for the onset of eruptive phenomena, including the
eruption of prominences, coronal mass ejections, and flares. It has
been formulated as a catastrophe or as an instability in the framework
of ideal magnetohydrodynamics (MHD).

The formulation as \emph{catastrophe} involves a sequence of
equilibria, i.e., the equilibrium manifold in parameter space, and an
``evolutionary scenario'' for the motion of the system point on the
manifold as a control parameter evolves continuously (representing
gradual changes at the boundary). Thus, it includes a model for the
pre-eruptive evolution. A catastrophe occurs if the system point
encounters a critical point on the equilibrium manifold. Most relevant
for solar eruptive phenomena is the case that the critical point is an
end point, or nose point, of the equilibrium manifold in the direction
of the changing parameter. The catastrophe then occurs by a \emph{loss
of equilibrium}, sometimes also referred to as
\emph{``non-equilibrium''}.

The formulation as \emph{instability} considers the evolution of a
small perturbation acting on an equilibrium at any point on the
equilibrium manifold. A full description of instability includes the
temporal evolution of the perturbation, but in order to find a
criterion for onset of eruption, only the point(s) of marginal
stability must be located in parameter space. As a parameter changes,
the system point moves from the stable part of the equilibrium
manifold across a point of marginal stability to the unstable part,
i.e., in this formulation the equilibrium is not lost but turns to an
unstable equilibrium. A model for the pre-eruptive evolution does not
enter here; the points of marginal stability are independent of the
pre-eruptive evolution.

The modeling of solar eruptions has so far mostly used either a
catastrophe formulation or an instability formulation, although they
are related to each other. An analysis of this relationship should be
helpful for unifying some of the independent developments in the
modeling, which we summarize next.

A model of eruption onset from the force-free equilibrium of a flux
rope was established by \citet{vanTend&Kuperus1978} who focused on
instability, but also related the instability to the fact that the
equilibrium may be lost \cite[see also][]{Molodenskii&Filippov1987}.
They considered a translationally invariant coronal current in the
center of a magnetic flux rope above a plane photospheric surface. The
current was approximated as a line current immersed in an external
poloidal field $B_\mathrm{e}$, and only its external, large-scale
equilibrium was analyzed. It was found that the height dependence
$B_\mathrm{e}(h)$ determines whether the configuration is stable or
unstable. The current is unstable to an upward displacement if
$B_\mathrm{e}$ decreases sufficiently rapidly with height $h$ above
the boundary surface. In the two-dimensional (2D) translationally
invariant geometry, the ``decay index'' $n=-d\ln B_\mathrm{e}/d\ln h$
must exceed $n_\mathrm{cr}=1$ for instability. This critical value was
derived under the assumption that any change of current produced by
the perturbation can be neglected, which is consistent with
conservation of magnetic flux between the current channel and the
boundary surface in the limit of vanishing current channel radius $a$
\citep{Forbes1990}. A slightly higher value results if the constraint
of flux conservation is imposed for $a>0$; then
$n_\mathrm{cr}=1+1/(2c)$, where $c=\ln(2h/a)+1$
\citep{Demoulin&Aulanier2010}.

An MHD description of the configuration, including internal force-free
equilibrium of the current channel, was developed by
\citet{Priest&Forbes1990} and \citet{Forbes&Isenberg1991} and further
elaborated in a series of papers by \citet{Isenberg&al1993},
\citet{Forbes&Priest1995}, \citet{Lin&Forbes2000}, and
\citet{Lin&vanBallegooijen2002}. All of these investigations described
the onset of eruption as the occurrence of a catastrophe. The
condition of flux conservation between the current channel and the
photosphere was adopted in some cases, but other assumptions were
considered as well, in order to model the changes in photospheric flux
budget (flux cancellation or emergence) which are often observed in
the pre-eruption phase \citep{Martin&al1985, Feynman&Martin1995}.
Various evolutionary scenarios and external field models were
analyzed. Accordingly, various locations of the critical point in
parameter space were obtained.

More recently, \citet{Longcope&Forbes2014} have found that a flux rope
in quadrupolar external field can reach a catastrophe along various
evolutionary paths, depending on the detailed form of the initial
equilibrium. Some equilibria can be driven to a catastrophe and
instability through reconnection at a lower, vertical current sheet, a
process often referred to as ``tether cutting'' \citep{Moore&al2001}.
While other equilibria can be driven to a catastrophe and instability
through reconnection at an upper, horizontal current sheet, a process
referred to as ``breakout'' \citep{Antiochos&al1999}. Some equilibria
can be destabilized by both processes, but others only by one and not
the other. Still other equilibria undergo no catastrophe and
instability, but evolve at an increasingly rapid rate in response to
slow steady driving.

The occurrence of a catastrophe has also been demonstrated for
toroidal current channels. \citet{Lin&al1998} considered a toroidal
flux rope encircling the Sun in the equatorial plane with an induced
current in the solar surface, or equivalently, an image inside the Sun
of the current channel. \citet{Lin&al2002} studied a toroidal current
channel one half of which is submerged below the (plane) photosphere.
In this geometry, the submerged half of the channel represents the
image current, but the evolution of the channel's major radius implies
that the footpoints move across the solar surface. The latter
unsatisfactory feature was remedied by \citet{Isenberg&Forbes2007};
however, the resulting complex expressions for line-tied equilibrium
of a partial torus have not yet allowed a determination of the
location of catastrophe or the onset of instability in general form.

The freely expanding toroidal current channel investigated in
\citet{Lin&al2002} is essentially a tokamak equilibrium \cite[or
Shafranov equilibrium,][]{Shafranov1966} whose external poloidal field
is due to a pair of point sources. This equilibrium was first
explicitly given in \citet{Titov&Demoulin1999}. The expansion
instability of the Shafranov equilibrium is referred to in fusion
research as one of the axisymmetric tokamak modes (the other one being
a rigid displacement along the axis of symmetry). Its first
consideration \citep{Osovets1959} gave the threshold for instability
as $n=-d\ln B_\mathrm{e}/d\ln R>n_\mathrm{cr}=3/2-(c-1)/[2c(c+1)]$,
where $c=\mathcal{L}/(\mu_0R)=\ln(8R/a)-2$, and $\mathcal{L}$, $R$,
and $a$ are the inductance and the major and minor radii of the torus,
respectively. The derivation used the large aspect ratio approximation
$R\gg a$ for the inductance $\mathcal{L}$, neglected the internal
inductance of the current channel, and assumed that the minor radius
does not change as the torus expands in a vacuum field. The term
$(c-1)/[2c(c+1)]<0.1$ for all $c>1$, so the threshold of instability
lies close to $3/2$. The instability was also considered by
\citet{Titov&Demoulin1999}, who estimated $n_\mathrm{cr}\sim2$, and by
\citet{Kliem&Torok2006}, who obtained $n_\mathrm{cr}=3/2-1/(4c)$,
assuming that the minor radius expands proportionally to the major
radius, and they called the instability a ``torus instability''; both
investigations were performed without awareness of the original work
by \citeauthor{Osovets1959}. An instability of this type was also
realized (without quantifying it) as a possible cause of eruptions by
\citet{Krall&al2000}. \citet{Olmedo&Zhang2010} proposed an analytical
model for the instability of a line-tied partial torus, and found
$n_\mathrm{cr}\to2$ in the limit of a full torus but surprisingly low
values for $n_\mathrm{cr}$ (even below unity) if one half or less of
the torus extends above the boundary. Numerical verifications of the
instability for line-tied partial tori found threshold values in the
range $n_\mathrm{cr}\approx1.5\mbox{--}2$ \citep{Torok&Kliem2007,
Fan&Gibson2007, Aulanier&al2010, Fan2010}.

\citet{Demoulin&Aulanier2010} extended the consideration of both
catastrophe and instability to arbitrary geometry of the current
channel, intermediate between linear and toroidal shapes. They
estimated that the instability threshold then typically falls in the
range $n_\mathrm{cr}\sim1.1\mbox{--}1.3$ and argued that catastrophe
and instability are ``compatible and complementary. In particular,
they agree on the position of the instability if no significant
current sheets are formed during the long-term evolution of the
magnetic configuration.'' Their arguments are based on the facts that
catastrophe and instability are related in general and that the
investigations cited above employed the same force balance determining
the external equilibrium of the current channel. This suggests that
torus instability (and its 2D variant) could possibly occur at the
critical point in these catastrophe models.

Here we perform a detailed consideration of the relationship between
catastrophe and instability in toroidal geometry, verifying that torus
instability is indeed the instability occurring at the catastrophe
studied by Priest, Forbes, Lin and co-workers. The catastrophe point
is located exactly at the major torus radius $R$ where
$n(R)=n_\mathrm{cr}$, for all cases considered. We also show a case in
which the change of a control parameter (i.e., a certain evolutionary
scenario) leads to neither a catastrophe nor an onset of instability.
However, another control parameter in this system does yield
catastrophic/unstable behavior.

For simplicity, we will use solar nomenclature in the following,
bearing in mind that the situation is generic for eruptions
originating in the low-density hot atmosphere of a magnetized, dense
star or accretion disk \citep{Yuan&al2009}. Similarly, we will use
``expansion'' of the current channel to represent any change of the
current channel's major radius in response to changes at the
photospheric boundary. Typically, expansion is observed prior to solar
eruptions, and the models considered here all exhibit expansion.

We present a discussion of the general relationship between
catastrophe and instability in Section~\ref{s:c+i}, introduce the
basic eruption model in Section~\ref{s:model}, and then study a number
of catastrophe scenarios in bipolar (Section~\ref{s:2p}) and
quadrupolar (Section~\ref{s:4p}) ambient field.
Section~\ref{s:conclusion} gives the conclusion.

\section{Catastrophe and instability}
\label{s:c+i}

\begin{figure*}[t]                                            
\centering
\includegraphics[height=6.5cm]{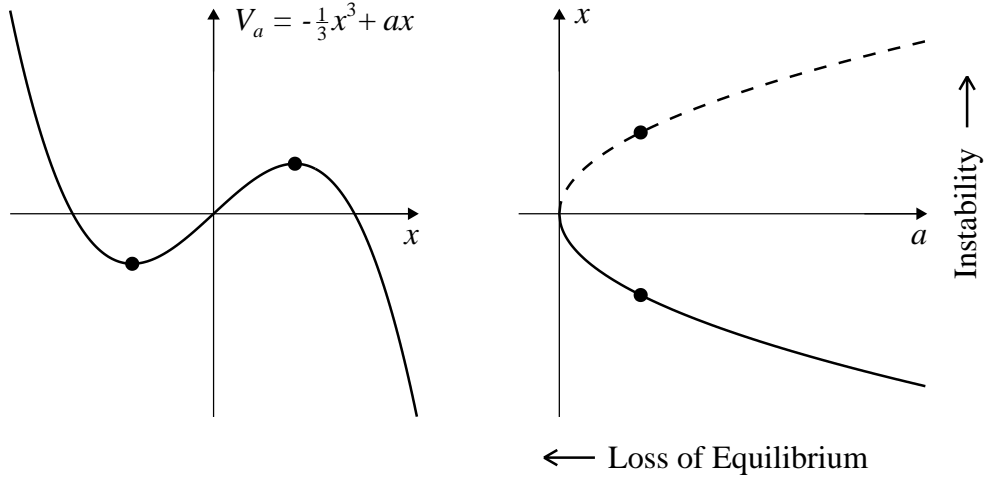} 
\caption{\label{f:fold}
 Fold catastrophe.
 \textit{(Left)} Potential $V_a(x)=-\frac{1}{3}x^3+ax$ for $a=1$.
 \textit{(Right)} Equilibrium manifold $x^2-a=0$; the stable
 (unstable) branch is plotted solid (dashed).
 The two equilibrium positions are marked in both plots.}
\end{figure*}

Catastrophe theory analyzes nonlinear systems that exhibit abrupt
changes of behavior, called catastrophes, and are governed, at least
locally in the vicinity of the point(s) of change, by a smooth
potential function $V_a(x)$ that depends upon at least one 
``behavior'' variable (or ``active'' variable) $x$ and at least one
``control parameter'' $a$. The force acting on the system in the space
of the behavior variable is given by $-\mathrm{d}V_a/\mathrm{d}x$, so
that the equilibrium positions are given by
$\mathrm{d}V_a/\mathrm{d}x=0$. Catastrophes occur where these are not
simple minima or maxima, but one or more higher derivatives of the
potential vanish as well at so-called degenerate critical points. The
simplest catastrophe thus occurs for a cubic potential with one
control parameter, $V_a=-\frac{1}{3}x^3+ax$, which has an inflexion
point at $x=a=0$. Figure~\ref{f:fold} (left panel) illustrates this
potential in the domain $a>0$, where it has a minimum (stable
equilibrium) at $x=-a^{1/2}$ and a maximum (unstable equilibrium) at
$x=a^{1/2}$. The equilibrium branches in the $a$-$x$ plane are plotted
in the right panel of Figure~\ref{f:fold}. As $a$ approaches zero, the
two extrema of the potential approach each other and disappear upon
merging in the inflexion point of the pure cubic function, which is an
end point of the pair of equilibrium branches. The catastrophe
occurring at $a=0$ is the \emph{fold catastrophe}. It occurs by a
\emph{loss of equilibrium}, since both equilibria are lost when the
control parameter $a$ is reduced below zero.

From the above it is obvious that every fold catastrophe must be
associated with an instability. The two equilibrium branches that join
at the catastrophe point are a continuous curve, and the catastrophe
point lies at the transition between the stable and unstable parts of
the curve, i.e., it is a point of marginal stability. For a system
evolving along a sequence of stable equilibria, both $x$ and $a$ may
be regarded as parameters of the equilibrium. For the toroidal current
channel studied below, the major radius $R$ is a natural choice for
the behavior variable, and one of the parameters specifying the
external poloidal field $B_\mathrm{e}$ is a natural choice for the
control parameter, for example, the strength of its sources, $q$, or
its decay index $n$. However, it is equally justified to regard $R$ as
a parameter describing the geometric properties of the equilibria. One
can consider an equilibrium sequence of toroidal current channels of
varying $R$, with fixed geometry of the sources of $B_\mathrm{e}$, and
compute the source strength $q_\mathrm{eq}(R)$ giving equilibrium for
each $R$. This is equivalent to following the equilibrium curve $a(x)$
in Figure~\ref{f:fold} by changing $x$. In this consideration, a loss
of equilibrium in the sense of catastrophe theory does \emph{not}
occur, but instability will set in as the  degenerate critical point
is crossed, resulting in an abrupt transition $x\to\infty$ that is
identical to the catastrophe occurring as $a$ is reduced below zero.

In most cases, the problem is not symmetric in $x$ and $a$. Often the
derivative $\partial V_a/\partial a$ does not represent any physical
quantity and is not related to the equilibrium positions of the
system. The latter is true in particular in catastrophe theory which
considers linear dependencies on the control parameters in the
vicinity of the degenerate critical points. Nevertheless, the
consideration of instability is not restricted to changes in the
control parameters, but analyzes in general how the change of any
variable describing the equilibrium affects its stability.

Since both the control parameter $a$ and the behavior variable $x$
change as the system point moves along the stable equilibrium branch
toward the point of catastrophe and instability, it is not trivial to
distinguish in a remote observation, like in the case of a solar
eruption, whether an equilibrium ceases to exist or goes unstable
\cite[see also the discussion in][]{Demoulin&Aulanier2010}. However,
by definition, it is a control parameter whose evolution causes the
system point to move along the stable equilibrium branch toward the
critical point. Typically, this can be the total flux of the external
field \citep{Bobra&al2008, YNSu&al2011, Savcheva&al2012c}, the
geometry of its sources which sets the height profile of the decay
index \citep{Torok&Kliem2007}, its shear whose increase causes a
magnetic arcade to expand and eventually collapse, forming a flux rope
\citep{Mikic&Linker1994}, or the twist of a flux rope rooted in a
rotating sunspot \citep{Amari&al1996a, Torok&Kliem2003, XLYan&al2012}.
The observations do not indicate that an external driver typically
operates directly at the height of current-carrying flux, although a
gradual increase of its footpoint separation may cause the flux to
ascend in some cases. In the vicinity of the critical point, a
fluctuation of any variable can cause the abrupt change of system
behavior.

\begin{figure*}[t]                                            
\centering
\includegraphics[height=6.5cm]{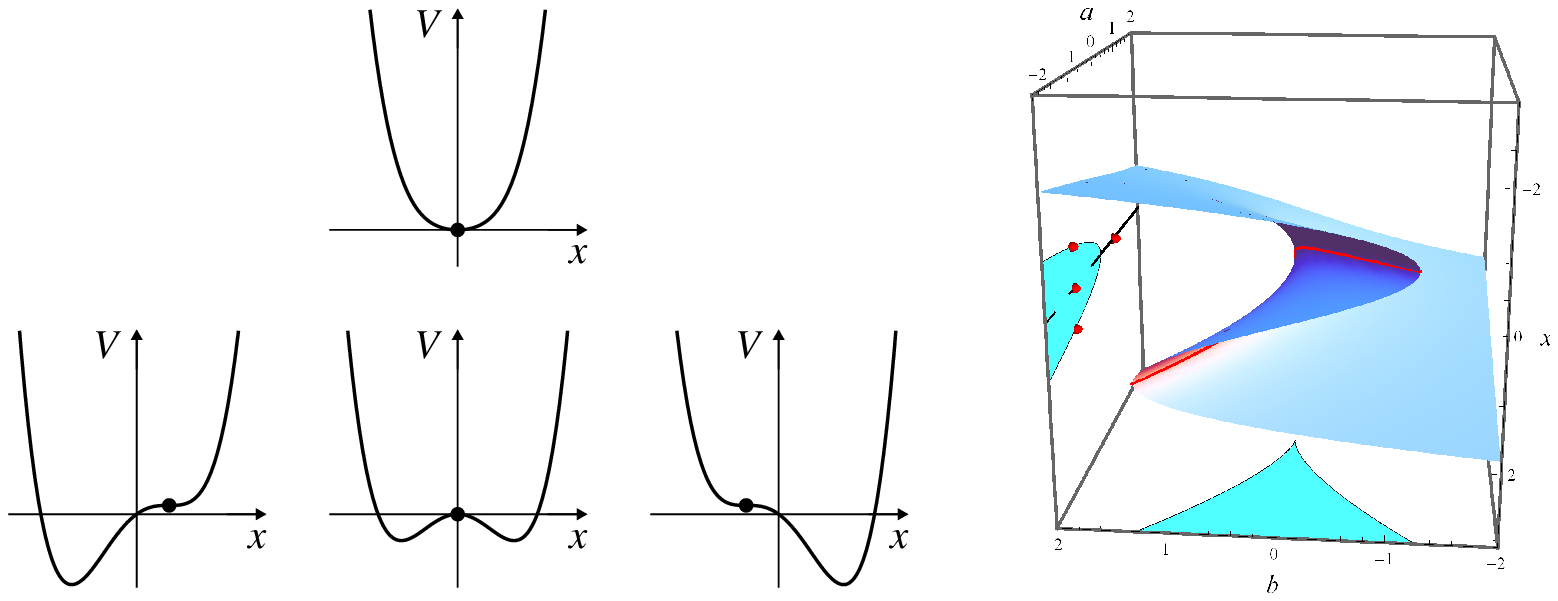}  
\caption{\label{f:cusp}
 Cusp catastrophe. 
 \textit{(Left four panels)} Potential
 $V_{ab}(x)=\frac{1}{4}x^4+\frac{1}{2}ax^2+bx$.
 Top plot: $a=0.6$, $b=0$. Bottom row: $a=-1.2$ and $b=2(-a/3)^{3/2}$,
 0, and $-2(-a/3)^{3/2}$ (from left to right).
 \textit{(Right panel)} Equilibrium manifold $x^3+ax+b=0$.
 The fold curve, $4a^3+27b^2=0$, is shown as a red line on the
 equilibrium manifold and also projected on the $b$--$a$ and $a$--$x$
 planes; the enclosed regions of unstable equilibria are colored in
 cyan. The line $x=b=0$ is added
 in the $a$--$x$ projection to complete the bifurcation diagram. The
 same four equilibrium positions are marked in the plots of the
 potential and bifurcation diagram.
 }
\end{figure*}

On the other hand, in a numerical simulation one has the freedom to
evolve a control parameter \cite[e.g.,][]{Chen&Shibata2000,
Amari&al2003, Torok&Kliem2003, Mackay&vanBallegooijen2006,
Aulanier&al2010, Torok&al2013} or to change the behavior variable
\cite[lifting a flux rope into the torus-unstable height range by a
prescribed velocity perturbation,][]{Fan&Gibson2007, Kliem&al2012}.
One can of course also test a configuration on any position of the
equilibrium manifold for its stability, independent of an evolutionary
scenario \cite[e.g.,][]{Lionello&al1998, Torok&al2004, Kliem&al2013}.

The next higher catastrophe arises with a potential
$V_{ab}(x)=\frac{1}{4}x^4+\frac{1}{2}ax^2+bx$. For $a>0$ this
potential has one minimum, but for $a<0$ there is a range of the
second parameter, $|b|<2(-a/3)^{3/2}$, inside which the potential has
two minima enclosing a maximum. Outside this range there is again only
one minimum (Figure~\ref{f:cusp}, left four panels). For $a<0$ this
maps to the well-known S-shaped equilibrium curve in $b$--$x$ space
which has three branches in the range $|b|<2(-a/3)^{3/2}$ and one
branch outside (Figure~\ref{f:cusp}, right panel). The nose points of
the equilibrium curve correspond to the merging of the maximum of the
potential with one of the minima in an inflexion point, i.e., they
represent fold catastrophes. Again, these are points of marginal
stability, where the unstable branch in the middle part of the
S-shaped equilibrium curve smoothly connects to a stable branch. Now,
if $a$ increases, approaching zero, the three extrema of the potential
approach each other. In $b$--$x$ space this corresponds to a shrinkage
of the S-shaped part of the equilibrium curve. The points of fold
catastrophe lie on two sections of a curve which approach each other
as $a$ increases.  As $a\to0$, all three extrema of the potential and
the two sections of the fold curve merge in the point of higher
degeneracy, $x=a=b=0$, where the \emph{cusp catastrophe} occurs. This
is a cusp point of the projected fold curve in the $b$--$a$ plane, but
the projection in the $a$--$x$ plane shows that the curve is smooth in
$b$--$a$--$x$ space. More generally, the equilibrium manifold, the
surface given by $x^3+ax+b=0$, is everywhere smooth, although it is
folded in $\{a<0\}$ (Figure~\ref{f:cusp}, right), since both
derivatives
$\partial b(x,a)/\partial x$ and $\partial b(x,a)/\partial a$ are
everywhere continuous.

With the exception of the cusp point, a loss of equilibrium occurs as
the system point crosses the line of fold catastrophes on a path lying
in the equilibrium manifold and  coming from the stable part of the
manifold. At the cusp point, the fold line can be crossed along a
smooth path that stays on the equilibrium manifold; coming from the
stable side, this must always occur by a change from $a\ge0$ to $a<0$
along the path. These two facts are perhaps most obvious from the
three-dimensional plot of the equilibrium surface and fold curve in 
Figure~\ref{f:cusp}, but they can also be seen in the $b$--$a$ plane of
the control parameters, where they represent crossings of the
projected fold curve in opposite directions. Once arrived on the
unstable part of the equilibrium manifold after passing through the
cusp point, any perturbation will cause the system to rapidly move to
one of the neighboring stable equilibrium positions, which is a
catastrophe although the move will only be a tiny one in practice,
and, of course, is an instability as well. Thus, the cusp catastrophe
does not occur by a loss of equilibrium, but by a change of the nature
of the equilibrium from stable to unstable. This evolutionary sequence
can be  termed a \emph{loss of stability}. The different types of
catastrophe are also obvious in the plots of the potential on the left
side of  Figure~\ref{f:cusp}. A loss of equilibrium occurs in the
horizontal transition from the middle panel to and beyond one of the
outer panels in the bottom row, and a loss of stability occurs in the
downward vertical transition between the middle panels.

Two important aspects must be noted for the relationship between
catastrophe and instability. First, instability is part of the cusp
catastrophe as this catastrophe occurs by the motion of the system to
the unstable part of the equilibrium manifold. Second, the term ``loss
of stability'' is \emph{not} synonymous with ``instability.''
\emph{Both} types of catastrophe---by loss of equilibrium and by loss
of stability---are associated with instability. The latter is
visualized by the (pitchfork) bifurcation diagram in the $a$--$x$ plane
(Figure~\ref{f:cusp}, right): here both the fold and cusp catastrophes
occur when the fold line is crossed from the stable to the unstable
part of the diagram.

We have seen that the cusp catastrophe occurs at a sub-manifold of the
manifold of fold catastrophes, which itself is a sub-manifold of the
equilibrium manifold. The dimensionality is reduced by one at each
level. This relationship extends to the manifolds of the higher
catastrophes, since the degenerate critical  points of a certain order
are always also degenerate critical points of lower order.
Consequently, the fold catastrophe is in general infinite times more
likely than any higher catastrophe.

The sample plots of the potential in Figure~\ref{f:cusp} also show
that a catastrophe can occur only by either a loss of equilibrium or a
loss of stability. In the first case the minimum disappears as the
slope $\mathrm{d}V/\mathrm{d}x$ changes sign only on one side, and in
the second case the slope changes sign on both sides simultaneously.
In the case of more than one behavior variable, this holds true for
each behavior variable and thus in general. Hence, any catastrophe is
related to an instability.

For the modeling of solar eruptions, the values of the control
parameters can vary in a wide range from event to event. Therefore, 
only the loss of equilibrium occurring in a fold catastrophe and the
associated instability are relevant in practice if the model contains
only one behavior variable. The occurrence of a higher catastrophe is
a special case, but any eruption mechanism must be able to operate in
a wide parameter range. Here it doesn't matter whether the loss of
equilibrium occurs in the potential of the fold catastrophe or in a
potential associated with one of the higher catastrophes.
Additionally, some of the higher catastrophes, like the cusp
catastrophe, do not provide a large change of the system.

If the model includes a second behavior variable, for example the
horizontal position of current-carrying flux, which may change in
response to asymmetric changes in the photospheric flux distribution,
then \emph{umbilic catastrophes} arise. \citet{Lin&al2001}
demonstrated this for a 2D flux rope equilibrium subject to flux
emerging only on one side of the rope. The potentials for the umbilic
catastrophes are at least cubic in at least one behavior variable.
Therefore, these catastrophes are sub-manifolds of the fold
catastrophe for at least one behavior variable \cite[see][Chapters~9.6--9.8
for detail]{Poston&Stewart1978}. It thus appears that the fold
catastrophe and its associated instability are most relevant in this
case as well.

\section{Model and basic equations}
\label{s:model}

We consider a self-similarly evolving toroidal current channel of
major radius $R$ and minor radius $a$ immersed in a given bipolar or
quadrupolar external field as our model for the source region of
eruptions. The current channel runs in the center of a toroidal flux
rope. Pressure and gravity are neglected, since the Lorentz force
dominates in strong active-region fields low in the corona, where most
major eruptions arise. While the model appears simplistic at first
glance, particularly in apparently missing the solar surface, it does
contain all the basic elements needed to describe a catastrophe or
instability of flux carrying a net current located above the
photospheric boundary:
(1) a realistic representation of the external poloidal field in
    bipolar and simple quadrupolar active regions;
(2) the flux rope of a prominence or filament
    channel; and
(3) the oppositely directed image current, given by the lower half of
    the torus.
Also see the discussion of the proper elements to be included in such
a model in \citet{Lin&al2002} and \citet{Demoulin&Aulanier2010} and
the support for the presence of net currents from recent
investigations of the current distribution in active regions
\citep{Ravindra&al2011, Georgoulis&al2012, Torok&al2014}. We also
neglect any external toroidal field components to facilitate an
analytical description. The simplicity of the model serves our aim to
determine the relation between catastrophe and torus instability.  The
model yields a transparent expression for the equilibrium manifold,
allowing us to consider a number of cases without mathematical
complexity, one of them fully analytically.

Essentially the same model was used in the consideration of the torus
instability by \citet{Osovets1959} and \citet{Kliem&Torok2006}, so
that we can directly refer to their results. For the purpose of
comparing catastrophe and instability, it is necessary that both are
described using the same or compatible approximations.

The model in its simplest form lacks photospheric line tying of the
flux and implies that the footpoints of the current channel move
across the solar surface. We demonstrate below for one of our cases
that a simple modeling of the line tying effect can be included and
does not change the result in this particular case. The motion of the
footpoints across the solar surface hardly affects the threshold of
instability, since only infinitesimally small changes of the major
radius are considered in determining the threshold. However, the
threshold does depend on the shape of the flux rope and on the
strength of the external toroidal (shear) field component, with our
choice of full toroidal shape (i.e., no line tying) and vanishing
external toroidal field giving a relatively low threshold value.

The system is governed by three equations which describe the external
equilibrium (i.e., the force balance in the major toroidal direction
at the toroidal axis), the internal force-free equilibrium of the
current channel (in the direction of the minor radius), and the
evolution of the flux enclosed by the torus as the major radius
changes.

The \emph{external equilibrium} of a toroidal current channel in a
low-beta plasma is known as the Shafranov, or tokamak, equilibrium
\citep{Shafranov1966, Bateman1978}. It is obtained from the following
force balance
\begin{eqnarray}                                               
\rho_m\frac{\mathrm{d}^2R}{\mathrm{d}t^2}
 &=&\frac{\mu_0I^2}{4\pi^2a^2R}
   \left[\ln\left(\frac{8R}{a}\right)-\frac{3}{2}+\frac{l_i}{2}\right]
  -\frac{IB_\mathrm{e}(R)}{\pi a^2}                        \nonumber\\
 &=&0\,,
\label{e:Shafranov}
\end{eqnarray}
where the first term describes the Lorentz self-force of the current
(also referred to as the hoop force) and the second term describes the
Lorentz force provided by the external poloidal field
$B_\mathrm{e}(R)$. In the present configuration, the hoop force
includes the repulsive force due to the image current. Here $\rho_m$
is the mass density in the torus, $I$ is the total ring current, and
$l_\mathrm{i}$ is the internal inductance per unit length of the ring.
$l_\mathrm{i}$ is of order unity if the radial profile of the current
density is not strongly peaked in the center of the torus, a situation
expected to be representative of the flux in solar filament channels,
and thus its specific value is only of minor influence on the
equilibrium. We adopt the value $l_\mathrm{i}=1$ as in
\citet{Lin&al1998, Lin&al2002}, valid for the linear force-free
equilibrium of a current channel \citep{Lundquist1951}, which is a
natural choice for a relaxed force-free equilibrium. The value 
$l_\mathrm{i}=1/2$ for a force-free equilibrium with uniform current
density, used in \citet{Kliem&Torok2006}, yields nearly the same
locations of the catastrophe and instability points for the cases
considered in Sections~\ref{s:2p} and \ref{s:4p}, which of course
also coincide in each case. The expression
in brackets in Equation~(\ref{e:Shafranov}) is exact for
large aspect ratio, $R\to\infty$. It remains a good approximation
(within 10 percent of the exact value) down to relatively moderate
aspect ratios of order $R/a\sim10$ and deviates from the exact value
by up to a factor $\approx2$ for lower aspect ratios
\citep{Zic&al2007}. The force balance (\ref{e:Shafranov}) yields an
equilibrium current
\begin{equation}                                               
I(R,a)=\frac{4\pi RB_\mathrm{e}(R)}{\mu_0c_1(R/a)}\,,
\label{e:ee}
\end{equation}
where $c_1(R/a)=\left[\ln(8R/a)-3/2+l_\mathrm{i}/2\right]$ has been
used as an abbreviation.

The \emph{internal equilibrium} of the current channel must be close
to a force-free field for the low plasma beta characteristic of source
regions for solar eruptions ($\beta\sim10^{-4}\mbox{--}10^{-2}$ in the
core of active regions). If a force-free field expands, it remains
force free if the expansion is self-similar. Therefore, assuming
\begin{equation}                                               
\frac{R}{a}=\mbox{const}
\label{e:ie}
\end{equation}
is a reasonable approximation for the gradual pre-eruptive evolution
of a single torus along a sequence of nearly force-free equilibria.
This is even more so because the expressions depend on the ratio $R/a$
only logarithmically. Numerical simulations of the torus instability
for small plasma beta indicate its initial evolution to be
approximately self-similar as well. The assumption of self-similarity
implies that the distribution of the current density in the cross
section of the current channel remains unchanged; it is thus
consistent with the assumption $l_i=\mbox{const}$, which has usually
been adopted in modeling the evolution toward solar eruptions, and
with the relation $aI(R,a)=\mbox{const}$, which has been used in
\citet{Lin&al2002} and other studies of the catastrophe.

It should be kept in mind  that self-similarity is not always a good
approximation. For the model of a flux rope encircling the Sun
considered in \citet{Lin&al1998}, it does not apply as long as the
major rope radius is comparable to the solar radius. While the rope
expands, its image contracts, which is opposite to self-similar
behavior of the system as a whole. This model behaves approximately
self-similar when $R\gg R_\odot$. Two-dimensional models that place
the source of the external field under the current channel, e.g., a
line dipole or quadrupole \citep{Forbes&Isenberg1991,
Isenberg&al1993}, are similar in this regard.

Finally, the equation governing the evolution of the \emph{poloidal
flux enclosed by the torus} yields an expression for $I(R,a)$. In the
solar case, the enclosed poloidal flux has two sources, namely
subphotospheric sources of the external poloidal flux and the coronal
current that provides the free magnetic energy for the eruption; they
are considered to be essentially independent of each other. The
sources of the external poloidal flux generally change in strength and
geometry on the long time-scale of the gradual pre-eruptive evolution,
with the former implying emergence or submergence of flux through the
photosphere. The flux in the corona adjusts to these slow changes
along a sequence of equilibria, which obey magnetostatic expressions.
The sources of the external flux do not change significantly on the
short time-scale of the eruption \citep{Schuck2010}, i.e., during the
development of instability. The coronal current generally changes both
in the equilibrium sequence and during the eruption, although its
subphotospheric roots tend to stay unchanged on the short time-scale
of the eruption. The conservation of frozen-in flux on the global
scale of the coronal current loop takes dominance over the conditions
at its footpoints in constraining the current.

We note that not all of the considerations above carry over to related
laboratory plasmas, the details depending on the specific setup. For
example, \citet{Osovets1959} considered a pulsed tokamak operation
with no external current drive. In this case, the current channel
expands and contracts in vacuum, and its current stems solely from
induction by the changing external poloidal flux generated in external
coils and linked by the torus.

The flux enclosed by the torus is $\Psi=\Psi_I+\Psi_\mathrm{e}$, where
the poloidal flux due to the ring current,
\begin{equation}                                               
\Psi_I(R,a)=\mathcal{L}(R,a)I(R,a)\,,
\label{e:Psi_I}
\end{equation}
is expressed in terms of the inductance of the torus, 
$\mathcal{L}(R,a)=\mu_0R\left[\ln(8R/a)-2+l_\mathrm{i}/2\right]$, and
the external poloidal flux is given by
\begin{equation}                                               
\Psi_\mathrm{e}(R)=-2\pi\int_0^RB_\mathrm{e}(r)\,r\,\mathrm{d}r\,.
\label{e:Psi_e}
\end{equation}
Here we have dropped the common factor $1/2$ referring to the upper
half of the enclosed flux (above the photosphere) and extended the
integral to $r=R$ instead of the accurate value $r=R-a=R(1-a_0/R_0)$,
where $R_0$ and $a_0$ are the initial values, to be consistent with
the treatment in \citet{Kliem&Torok2006}. This approximation
simplifies the resulting algebraic expressions. The neglect of the
factor $(1-a_0/R_0)$ causes only small quantitative changes in the
large aspect ratio approximation underlying
Equation~(\ref{e:Shafranov}). It could easily be incorporated in the
resulting expressions, without changing the qualitative results.

The evolution of the enclosed flux during changes of the major torus
radius depends on the occurrence of reconnection. We first consider
a case in which the field under the current channel has an X-type
magnetic configuration. This is a two-dimensional X-point field if the
external field is purely poloidal,
an X-line in the 3D view of our axisymmetric model. (The X-line
becomes a separator field line if an external toroidal field component
is present, and it coincides or approximately coincides with a
quasi-separator line running within a hyperbolic flux tube if the
photospheric boundary is also taken into account,
\citealt{Priest&Demoulin1995, Titov&al2002}). Expansion of the current
channel in the presence of an X-type structure is likely to be
associated with reconnection, both before the eruption
\cite[e.g.,][]{Aulanier&al2010} and during the eruption
\cite[e.g.,][]{Torok&Kliem2005}. The X-line acts as a seed for the
formation of a vertical current sheet and the onset of reconnection.
Since the time-scale of reconnection in the corona is shorter than the
time-scale of photospheric driving, typically reconnection acts
efficiently at the X-line and a large-scale vertical current sheet
does not develop in the evolution of a system on the equilibrium
manifold (different from the fast evolution during eruption). In the
rest of the volume the flux is assumed to be frozen in. Reconnection
under the current channel in this \emph{non-ideal MHD case} adds equal
amounts of positive and negative poloidal flux to the area between the
current and the photosphere. It also allows the flux rope to ``slide
through'' the external poloidal field: the amount of originally
overlying flux transferred by reconnection to the flux rope equals the
amount of flux added below the X-line.

Since the flux rope slides through the external field in the non-ideal
case, to a first approximation the functional form of
$B_\mathrm{e}(R)$ remains invariant as the major torus radius changes.
This approximation is supported by the agreement of the resulting
threshold value with many numerical and observational studies of the
torus instability. Gradual changes of the external field can thus be
described by changes of the parameters, $p$, in a given function
$B_\mathrm{e}(R,p)$.

In determining the instability threshold, the parameters of the
external field are treated as given. The enclosed flux is then
conserved in the non-ideal case,
\begin{eqnarray}                                               
\Psi(R,a)
  &=&\mathcal{L}(R,a)I(R,a)+\Psi_\mathrm{e}(R)     \nonumber\\
  &=&\Psi_0\,,
\label{e:fc_recon}
\end{eqnarray}
where $\Psi_0=\Psi(R_0,a_0)$. Here and in the following we use the
subscript 0 to denote initial values (of a reference equilibrium at an
arbitrary point on the stable part of the equilibrium manifold in
parameter space). 

As the parameters of the external field $B_\mathrm{e}(R,p)$ change in
the pre-eruptive evolution considered in the description of
catastrophe, both the force $IB_\mathrm{e}$ in
Equation~(\ref{e:Shafranov}) and the external flux $\Psi_\mathrm{e}$
given by Equation~(\ref{e:Psi_e}) change. Including the change
$\Delta\Psi_\mathrm{e}$ in the equation for the enclosed flux yields
\begin{eqnarray}                                           
\Psi(R,a,p)&=&\mathcal{L}(R,a)I(R,a)+\Psi_\mathrm{e}(R,p)  \nonumber\\
           &=&\Psi_0+\Delta\Psi_\mathrm{e}(R_0,p)\,, \label{e:fc+source}\\
\Delta\Psi_\mathrm{e}(R_0,p)
           &=&-2\pi\int_0^{R_0}
              [B_\mathrm{e}(r,p)-B_\mathrm{e}(r,p_0)]\,r\,\mathrm{d}r\,.
                                                          \label{e:source}
\end{eqnarray}
If the sources of the external field change in strength on the Sun,
flux must emerge or submerge through the photosphere. This is
represented by the term $\Delta\Psi_\mathrm{e}$. In our model with
toroidal symmetry, where the sources of $B_\mathrm{e}$ must be
symmetric with respect to the photosphere, this can still be thought
of as a gain or loss of flux through the photosphere. On the other
hand, if the photospheric flux distribution is rearranged with a fixed
strength of its sources, then the change is not associated with
emergence or submergence of flux for the frozen-in conditions on the
Sun. However, in the parametric representation of geometric changes
with fixed functional form $B_\mathrm{e}(R,p)$ in the present
formulation, the enclosed flux generally changes as $p$ changes. The
flux is exchanged between the area enclosed by the torus and the
exterior area in this case, not through the photosphere (this is
obvious from considering a varying distance of the sources from the
plane of the torus). As the torus slides through the external field in
response to a change of $B_\mathrm{e}$, it regains part or all of the
flux exchanged between the two areas. In particular, if the sources of
$B_\mathrm{e}$ are simply moved along the symmetry axis of the torus,
the torus radius can change proportionally to keep equilibrium, and in
this case the enclosed external flux stays invariant. Therefore, we
choose to use Equation~(\ref{e:fc_recon}) for the enclosed flux if $p$
describes the geometry of the sources of $B_\mathrm{e}$, and
Equation~(\ref{e:fc+source}) if $p$ describes their strength. The
resulting differences in the location of the catastrophe point remain
minor in large parts of parameter space (e.g., in the cases displayed
in Figures~\ref{f:2p_sigma}, \ref{f:4p_sigma}, \ref{f:4p_delta}, and
\ref{f:4p_nu} below) but they can be considerable for some parameter
combinations (the case shown in Figure~\ref{f:4p_delta2} is an
example).

If there is no X-line but rather a bald-patch separatrix surface
below the current-carrying flux, then a vertical current sheet cannot
immediately form if the current channel expands; it will form only
after a considerable expansion has led to  sufficient horizontal
constriction of the flux below the channel. For thin channels this
occurs relatively early \cite[e.g.,][]{Forbes&Isenberg1991,
Lin&al2002}, so that the remaining evolution on the equilibrium
manifold can be described using Equation~(\ref{e:fc_recon}) or
(\ref{e:fc+source}), but for thick channels this does not occur before
the eruption develops strongly \cite[e.g.,][]{Gibson&Fan2006}. In the
resulting absence of reconnection in the pre-eruptive evolution, both
parts of the enclosed flux are conserved individually, giving us the
simple equation for the case of ideal-MHD evolution
\begin{equation}                                               
\Psi_I(R,a)=\mathcal{L}(R,a)I(R,a)=\Psi_{I0}\,.
\label{e:fc_ideal}
\end{equation}
Since the torus cannot ``slide through'' the external field in ideal
MHD, the functional form of $B_\mathrm{e}(R)$ must change in this case
if the flux rope expands. This generally also includes the formation
of currents in the ambient volume.

Topological considerations of active-region evolution suggest that
either case can be realized \citep{Titov&Demoulin1999}. Both
possibilities were also supported by data analysis
\cite[e.g.,][]{Green&Kliem2009, Green&al2011}, active-region modeling
\cite[e.g.,][]{YNSu&al2011, Savcheva&al2012a}, and numerical simulation
\cite[e.g.,][]{Gibson&Fan2006, Aulanier&al2010}.

It is worth noting that Equations~(\ref{e:fc_recon}) and
(\ref{e:fc+source}) do not describe evolution in a vacuum, although
they are based on the assumption that the flux rope moves through the
external poloidal field. If the flux rope expanded in vacuum,
$\Psi_I(R)$ would be conserved (Equation~\ref{e:fc_ideal}), while
$\Psi_\mathrm{e}(R)$ would change according to
Equation~(\ref{e:Psi_e}) with the functional form of
$B_\mathrm{e}(R,p)$ being preserved exactly. Although the description
of the evolving external field as a parameter dependence of
$B_\mathrm{e}(R,p)$ with fixed functional form still contains an
element of vacuum behavior, it represents a reasonable approximation
to the MHD behavior of the system, as discussed above in relation to
Equation~(\ref{e:fc+source}).

Moreover, the modeling approach laid out above and also employed in
\citet{Kliem&Torok2006} should not be categorized as a ``circuit
model'', since it does not contain any element of an electric circuit.
There are no current sources or sinks. Rather the current is a
secondary quantity depending on the evolution of the magnetic flux,
and given by Equation~(\ref{e:Shafranov}), combined with
Equation~(\ref{e:fc_recon}), (\ref{e:fc+source}) or (\ref{e:fc_ideal})
which express MHD considerations. Similarly, the assumption made for
the internal equilibrium (Equation~(\ref{e:ie})) expresses a property of
a force-free field.

In the following, we consider only the non-ideal case, since a
reliable analytical description of the changing function
$B_\mathrm{e}(R)$ does not yet exist for the ideal-MHD case.
\citet{Kliem&Torok2006} have formally derived a torus instability
threshold $n_\mathrm{cr}=2$ in this case, using the parameterized form
$B_\mathrm{e}(R)=\hat{B}R^{-n}$, where $n$ is not prescribed but
determined from the condition of marginal stability; however, they
noted that the formulation was not self-consistent. A closer
consideration of this case suggests a completely different
description, focusing on the Lorentz forces formed in the ambient
flux. If the current channel expands self-similarly in the ideal-MHD
case, then the frozen-in field component $B_\mathrm{e}(R)$ within the
channel decreases proportionally to $R^{-2}$. Since $I\propto R^{-1}$
in this case (from Equations~(\ref{e:fc_ideal}) and (\ref{e:ie})), the
external force balance (\ref{e:Shafranov}) is not affected, i.e., in
this approximation no force resisting or amplifying the expansion is
induced within the channel. Numerical simulations of this case in the
zero-beta limit indicate that the expansion indeed tends to be
approximately self-similar (a result of force-freeness). The expansion
piles up the ambient flux above the current channel, creating a
downward-directed magnetic pressure gradient in the ambient flux.
Below the current channel, the flux is stretched upward, reducing the
curvature radius of the upward concave field lines, which creates an
upward Lorentz force. The global force balance tends to be dominated
by the opposing Lorentz forces created in the compressed or stretched
ambient flux. Their ratio, and hence the stability of the current
channel, again depends on the decay index of the external field, but
also quite significantly on the aspect ratio $R_0/a_0$, with thicker
tori being more stable. Numerical simulations indicate a threshold of
instability closer to the canonical value of 3/2 for moderate aspect
ratio but rising even above 2 for very thick tori; these will be
reported in a future study.

The flux equation (\ref{e:fc+source}) yields the following expression
for the current
\begin{equation}                                              
I(R,a,p)
 =\frac{ \mathcal{L}_0I_0+\Psi_\mathrm{e0}-\Psi_\mathrm{e}(R,p)
        +\Delta\Psi_\mathrm{e}(R_0,p)}
       {\mathcal{L}(R,a)}\,.
\label{e:I}
\end{equation}
The cases that are instead described by Equation~(\ref{e:fc_recon})
are covered by this expression if the term $\Delta\Psi_\mathrm{e}$ is
dropped. Inserting Equation~(\ref{e:ee}) into Equation~(\ref{e:I}) to
eliminate the current, and using Equation~(\ref{e:ie}), we obtain the
expression for the equilibrium manifold in the non-ideal case,
\begin{eqnarray}                                              
0&=&R^2B_\mathrm{e}(R,p)-R_0^2B_\mathrm{e0}            \nonumber\\
 && +\frac{c_1}{4\pi c_2}
   [\Psi_\mathrm{e}(R,p)-\Psi_\mathrm{e0}-\Delta\Psi_\mathrm{e}(R_0,p)]\,,
\label{e:manifold}
\end{eqnarray}
where the abbreviation
$c_2(R/a)=\mathcal{L}(R,a)/(\mu_0R)=c_1-1/2$ has been introduced.
Since the geometry of the flux rope is assumed to be
invariant in our model, the expression for the equilibrium manifold
shows in a particularly transparent form that the properties of the
equilibria are determined by the external field. For the consideration
of catastrophe and instability in our model, we do not need to compute
the whole field (although it is well known for the specific choices of
$B_\mathrm{e}(R,p)$ treated below).

To find the point of marginal stability for this model,
\citet{Kliem&Torok2006} determined at which radius $R$ the force
(Equation~(\ref{e:Shafranov})) resulting from an infinitesimal
perturbation of the equilibrium changes sign,
\begin{equation}                                              
\frac{\mathrm{d}}{\mathrm{d}R}
  \left.\left(
     \rho_m\frac{\mathrm{d}^2R}{\mathrm{d}t^2}\right)\right|_{R=R_\mathrm{eq}}
=0\,,
\label{e:instab}
\end{equation}
where $R_\mathrm{eq}$ is a radius on the equilibrium manifold
satisfying
$\left.\mathrm{d}^2R/\mathrm{d}t^2\right|_{R=R_\mathrm{eq}}=0$. In
their treatment Equations~(\ref{e:I}) and (\ref{e:ie}) were inserted
in (\ref{e:Shafranov}), which is equivalent to
Equation~(\ref{e:manifold}), and then the derivative was taken. The
resulting torus instability threshold is
\begin{equation}                                              
n_\mathrm{cr}=\frac{3}{2}-\frac{1}{4c_2}\,.
\label{e:n_cr}
\end{equation}
This can readily be verified to hold for \emph{any smooth function}
$B_\mathrm{e}(R,p)$ by taking the derivative of expression
(\ref{e:manifold}) for the equilibrium manifold (which immediately
also yields the threshold $n_\mathrm{cr}=2$ if the terms proportional
to $c_1/c_2$ are dropped). For aspect ratios in the range typically
considered (i.e., $R/a=3\mbox{--}100$), the second term in
Equation~(\ref{e:n_cr}) is a small number in the range 0.05--0.15.

Equation~(\ref{e:instab}) explicitly demonstrates that instability and
fold catastrophe are equivalent descriptions of the transition to a
non-equilibrium state. At the point of marginal stability, the force
resulting from an infinitesimal deviation from equilibrium changes
from a restoring to an amplifying force (vanishing derivative of the
left-hand side of Equation~(\ref{e:Shafranov})). This coincides with a
degenerate critical point of the underlying potential (vanishing
derivative of the expression for the equilibrium manifold in the
middle part of Equation~(\ref{e:Shafranov}), as discussed in
Section~\ref{s:c+i}), i.e., with a point of catastrophe. Thus any
catastrophe occurring in the expansion of the torus in the model
considered in this paper must occur at the threshold of torus
instability.

This also resolves an apparent problem indicated by the different
expressions for the enclosed flux in the description of catastrophe
(including the term $\Delta\Psi_\mathrm{e}$ in some cases) and
instability (excluding the term $\Delta\Psi_\mathrm{e}$). Since
$\Delta\Psi_\mathrm{e}(R_0,p)$ does not contribute to the derivative
of the equilibrium manifold (\ref{e:manifold}), the torus instability
threshold (\ref{e:n_cr}) is independent of its inclusion, i.e., the
two approximations are compatible with each other. Although the
position of the degenerate critical point in $R$--$p$ space depends on
whether or not $\Delta\Psi_\mathrm{e}(R_0,p)$ is included, it
coincides with the instability threshold (\ref{e:n_cr}) in either
case.

For consistency of the presentation, we will use an aspect ratio
$R/a=10$ in all applications that follow, so that the Lorentz
self-force of the current channel is well approximated by Shafranov's
expression in Equation~(\ref{e:Shafranov}). It should be noted that
the considerations above, in particular the
expression~(\ref{e:manifold}) for the equilibrium manifold, remain
valid for smaller aspect ratio because the inductance, and hence the
Lorentz self-force, then still depend on $R$ and $a$ in the same form
as Shafranov's expressions \citep{Zic&al2007}. Only the definition of
the numerical coefficients $c_1$ and $c_2$ differs.

The relatively high aspect ratio, in combination with the assumption
that half of the torus extends above the photosphere, implies a high
value of the twist. The field line pitch (the axial length for one
winding about the axis) in a force-free current channel is comparable
to the radial length-scale of the field, $a$. Therefore, a high twist
is unavoidable for high aspect ratio. We disregard the resulting
susceptibility of the current channel to helical kinking
\citep{Hood&Priest1979} and focus exclusively on the stability
properties with respect to toroidal expansion (a form of lateral
kinking), since it is this instability which is related to the
catastrophes investigated previously. A simultaneous consideration of
both instabilities in the framework of catastrophe theory (an umbilic
catastrophe) has, to our knowledge, not yet been performed. The
observations of filaments and prominences indicate that flux ropes in
the solar corona typically have a smaller aspect ratio and are stable
against the helical kink mode in the majority of cases.

\section{Catastrophe versus instability in a bipole field}
\label{s:2p}

\begin{figure*}[t]                                            
\centering
\includegraphics[angle=-90,width=.7\linewidth]{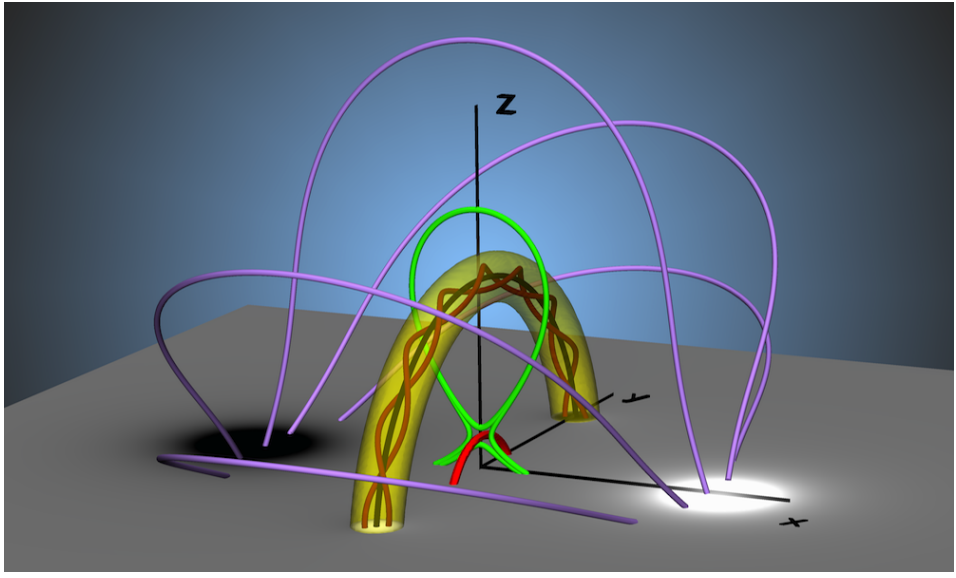} 
\caption{\label{f:2p_sigma_fl}
 Bipolar active-region model \citep{Titov&Demoulin1999} showing
 $B_z(x,y)$ in the bottom plane (saturated gray scale), the current
 channel of major radius $R$ and minor radius $a=0.1R$ as a
 transparent isosurface of current density, field lines of the
 force-free field in the current channel, field lines of the purely
 poloidal field external to the current channel considered in this
 paper, and the toroidal X-line as a red ring. The bottom plane is
 positioned at $z\approx0.2R$. The locations of the peak $|B_z|$
 values in the bottom plane indicate the positions of the sources of
 the external poloidal field at $\bm{x}=(\pm L,0,0)$. The major torus
 radius in this and the subsequent field line plots lies on the stable
 equilibrium branch close to the catastrophe point.}
\end{figure*}

We first consider a bipole as the source of an external field, with
the poles of strength $\pm q$ located at the symmetry axis of the
torus at distances $\pm L$ from the torus plane. This is identical to
the Titov--D\'emoulin model of an active region
\citep{Titov&Demoulin1999}, which has been successfully used in
qualitative and quantitative numerical modeling of a wide range of
solar eruptions \cite[e.g.,][]{Torok&Kliem2005, Schrijver&al2008b,
Kliem&al2012}. The external field in the torus plane is perpendicular
to the plane and given by
\begin{equation}                                              
B_\mathrm{e}(R)=\frac{\mu_0}{2\pi}\frac{qL}{(R^2+L^2)^{3/2}}\,.
\label{e:2p_Be}
\end{equation}
Such a configuration allows us to consider both scenarios for a
catastrophe considered previously in the context of solar eruptions,
i.e., changing the field amplitude \citep{Forbes&Isenberg1991,
Isenberg&al1993, Lin&al1998, Lin&vanBallegooijen2002}, here
parameterized by $q$, and changing the spatial scale of the field
\citep{Forbes&Priest1995, Lin&vanBallegooijen2002}, here parameterized
by $L$. The catastrophe for this system has already been investigated
in \citet{Lin&al2002}, using the more general approximation
$aI(R,a)=\mbox{const}$ in place of Equation~(\ref{e:ie}). For
comparison with the torus instability threshold, we repeat the
analysis here using Equation~(\ref{e:ie}) as in
\citet{Kliem&Torok2006}.

The decay index of the bipole field in the plane of the torus is
\begin{equation}                                              
n_\mathrm{bp}=-\frac{d\ln{B_\mathrm{e}}}{d\ln{R}}=3(L^2/R^2+1)^{-1}\,.
\label{e:2p_n}
\end{equation}
The torus instability threshold in the non-ideal MHD case
(Equation~(\ref{e:n_cr})) lies here at $R/L=[(6c_2-1)/(6c_2+1)]^{1/2}$,
i.e., slightly below unity. In terms of $\xi=R/L$, the expressions
required in Equation~(\ref{e:manifold}) are
\begin{equation}                                              
R^2B_\mathrm{e}(R)=\frac{\mu_0}{2\pi}\frac{q\xi^2}{(\xi^2+1)^{3/2}}
\label{e:2p_R^2Be}
\end{equation}
and
\begin{equation}                                              
\Psi_\mathrm{e}(R)=\mu_0q\left[\frac{1}{(\xi^2+1)^{1/2}}-1\right]\,.
\label{e:2p_Psie}
\end{equation}

To see whether and where the torus in the bipole field exhibits
catastrophic behavior, we choose a reference equilibrium in the stable
part of parameter space, i.e., $R=R_0<[(6c_2-1)/(6c_2+1)]^{1/2}L_0$
and $q=q_0$, and vary either the bipole strength as
$q(t)=\sigma(t)q_0$ with fixed geometry $L=L_0$, or the geometry as
$L(t)=\lambda(t)L_0$ with fixed bipole strength $q=q_0$. In the former
case, the torus must expand to find a new equilibrium if $\sigma$
decreases (which represents flux cancellation under the flux rope or a
general decay of an active region). Since $L$ is kept fixed, this
implies that the new equilibrium is situated at a radius with a
steeper slope for the external field, thus approaching the threshold
of the torus instability. If torus instability and catastrophe are
equivalent, a catastrophe must then occur. In the latter case, the
field strength at the original torus position decreases if $L$
increases (corresponding, for example, to active-region dispersal), so
that the torus must also expand to find a new equilibrium. Since the
equilibrium manifold depends on $R$ and $L$ only in the combination
$\xi=R/L$ (Equations~(\ref{e:manifold}), (\ref{e:2p_R^2Be}), and
(\ref{e:2p_Psie})), $R$ increases proportionally to $L$, representing a
simple rescaling of the configuration without approaching a loss of
equilibrium or the torus instability threshold (see
\citeauthor{Lin&al2002} \citeyear{Lin&al2002} and
Section~\ref{ss:2p_L} below).

\subsection{Changing the Source Strength}
\label{ss:2p_q}

\begin{figure}[t]                                             
\centering
\includegraphics[width=\columnwidth]{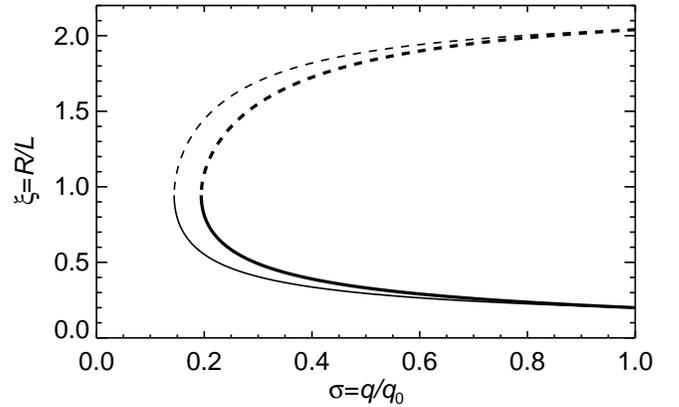} 
\caption{\label{f:2p_sigma}
 Equilibrium torus radius $R/L$ as a function of bipole strength
 $\sigma$ with the term $\Delta\Psi_\mathrm{e}$ in
 Equation~(\ref{e:manifold}) included (thick line) and excluded
 (thin line) for an aspect ratio of the current channel of $R/a=10$.
 Solid (dashed) lines represent stable (unstable) equilibria in
 this and all subsequent plots of the equilibrium manifold.}
\end{figure}

We set $q=\sigma(t)q_0$ and $L=L_0$. Inserting the expressions
(\ref{e:2p_R^2Be}) and (\ref{e:2p_Psie}) into
Equation~(\ref{e:manifold}) immediately yields an explicit expression
for the equilibrium curve $\sigma=f(R,R_0/L,R_0/a_0)$,
\begin{equation}                                              
\sigma=\frac{ 2\xi_0^2(\xi_0^2+1)^{-3/2}}
            {\begin{array}{l}
             2\xi^2  (\xi^2  +1)^{-3/2}                            \\
     \quad {}-(c_1/c_2)[(\xi_0^2+1)^{-1/2}-(\xi^2+1)^{-1/2}]
             \end{array}}\,.
\label{e:2p_sig_recon}
\end{equation}
Here only the denominator depends on $\xi$. It is straightforward to
verify that it has a maximum at $\xi=[(6c_2-1)/(6c_2+1)]^{1/2}$, which
is a minimum of the function $\sigma(R)$, i.e., the location of a fold
catastrophe point (a nose point of $R(\sigma)$). Inserting this
critical radius in expression~(\ref{e:2p_n}), the decay index of the
bipole field at the catastrophe point is found to be
$n_\mathrm{bp}=3/2-1/(4c_2)$---exactly the instability threshold given
in Equation~(\ref{e:n_cr})---, which verifies the correspondence
between catastrophe and torus instability of the flux rope.
Figure~\ref{f:2p_sigma_fl} illustrates the equilibrium and
Figure~\ref{f:2p_sigma} plots the function $R(\sigma)$ for $R/a=10$
($c_2=2.88$). The figure also shows a plot of the equilibrium manifold
obtained if Equation~(\ref{e:fc_recon}) is used instead of
Equation~(\ref{e:fc+source}). Catastrophe then occurs at the same
torus radius (same decay index) but at a somewhat different value of
the control parameter.

For comparison, \citet{Lin&al2002} find a catastrophe occurs at
$R/L=0.94$. Using their value for the aspect ratio $R_0/a_0=100$, our
expressions locate the catastrophe/instability point at nearly the
same value, $R/L=0.97$. \citet{Lin&al2002} use the force balance
(\ref{e:Shafranov}) for the external equilibrium, $I\propto a^{-1}$
for the internal equilibrium, and (for this result) the conservation
of flux according to Equation~(\ref{e:fc_recon}). The close agreement 
of the results indicates that the assumption (\ref{e:ie}), which
considerably simplifies the expressions for the equilibrium manifold,
is appropriate for our system.

\subsection{Changing the Length-scale}
\label{ss:2p_L}

Setting $q=q_0$ and varying $L(t)=\lambda(t)L_0$,
Equation~(\ref{e:manifold}), with the term $\Delta\Psi_\mathrm{e}$
dropped, becomes
\begin{eqnarray}                                              
&&  \frac{2\lambda^{-2}\xi^2}{(\lambda^{-2}\xi^2+1)^{3/2}}
   -\frac{c_1}{c_2}\left[1-\frac{1}{(\lambda^{-2}\xi^2+1)^{1/2}}\right]
                                                          \nonumber\\*
&=& \frac{2\xi_0^2}{(\xi_0^2+1)^{3/2}}
   -\frac{c_1}{c_2}\left[1-\frac{1}{(\xi_0^2+1)^{1/2}}\right]\,,
\label{e:2p_lam_recon}
\end{eqnarray}
where $\xi=R/L_0$ and $\xi_0=R_0/L_0$. One immediately sees that this
depends on $\lambda$ and $\xi$ only in the combination $\xi/\lambda$,
so that the equilibrium sequence $\xi(\lambda)=\lambda\xi_0$
represents a simple rescaling of the configuration, as discussed above
and first demonstrated in \citet{Lin&al2002}. Thus, the length-scale
$L$ of the bipole is not an appropriate control parameter to obtain
catastrophic or unstable behavior of the model.

The simple scaling relationship between $\lambda$ and $\xi$ breaks
down if photospheric line tying is included. We have attempted to
model this by employing the approximation for the inductance of a
line-tied current channel
\begin{equation}
\mathcal{L}(R)
   =\mu_0R\left[\frac{1}{2}\left( \ln\frac{8R}{a_\mathrm{f}}
                                 +\ln\frac{8R}{a_\mathrm{a}}\right)
                -2+\frac{l_\mathrm{i}}{2}\right]             \nonumber
\label{e:L_Garren}
\end{equation}
developed by \citet{Garren&Chen1994}. Here $a_\mathrm{f}$ and
$a_\mathrm{a}$ are the minor torus radii at the footpoints and apex of
the current channel, respectively. Setting $a_\mathrm{f}=a_0$ and
using Equation~(\ref{e:ie}) for $a_\mathrm{a}$, the above average
yields the additive correction $\ln(R/R_0)/2$ to the logarithmic term
in the inductance of a freely expanding torus, which must be applied
to the logarithmic term in the force balance (\ref{e:Shafranov}) as
well. The coefficients $c_1$ and $c_2$ are now functions of $R$ (or
$\xi$) but not of $\lambda$. However, since the correction is at most
moderate (due to the logarithmic dependence on $R$), since it is
applied to both coefficients, and since only the ratio $c_1/c_2$
enters the equations, the effect on the equilibrium curve $R(\lambda)$
remains very minor, so that a catastrophe still does not occur. This
appears quite plausible, since line tying tends to hinder the
expansion of the current channel in comparison with free expansion, so
that it is more difficult for it to expand into the torus-unstable
range as $L(t)/L_0$ increases.

\section{Catastrophe versus instability in the field of a linear
         quadrupole}
\label{s:4p}

\begin{figure*}[t]                                            
\centering
\includegraphics[angle=-90,width=.7\linewidth]{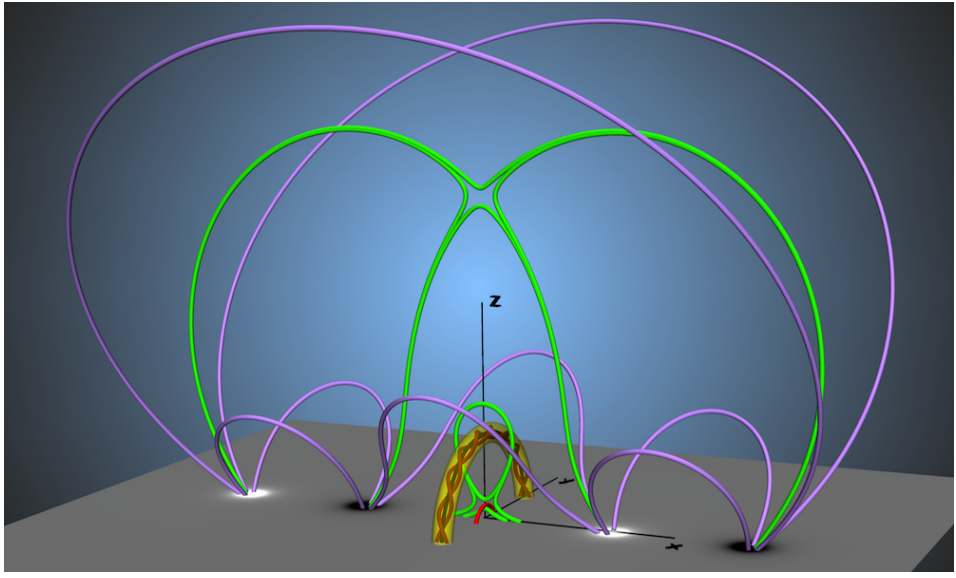} 
\caption{\label{f:4p_sigma_fl}
 Quadrupolar active-region model (generalized Titov-D\'emoulin
 equilibrium) shown in a format similar to Figure~\ref{f:2p_sigma_fl}.
 The bottom plane is here positioned at $z=0.1R$ to include the
 low-lying X-line in the display. The values of the parameters $R/a$,
 $\kappa$, and $\epsilon$ are identical to Figure~\ref{f:4p_sigma}.}
\end{figure*}

As a second realization of our model we consider the expansion of a
torus in the field of a linear quadrupole consisting of two nested
bipoles (denoted by subscripts 1 and 2); both are placed symmetrically
with respect to the torus plane at the symmetry axis of the torus.
This field can have a steeper slope than that of a single bipole,
especially below a magnetic null line (X-line), which is present for
a wide range of parameter combinations $(q_2/q_1, L_2/L_1)$ if the two
bipoles are oppositely directed (see Figure~\ref{f:4p_sigma_fl} for
an illustration). Thus, torus instability tends to
occur at a smaller $R$, and the catastrophe has also been found to
occur at a small height above the photosphere if the external field is
quadrupolar \citep{Isenberg&al1993}. All configurations considered
below include an X-line above and a second X-line below the current
channel for parameters in the vicinity of the catastrophe point,
except the configuration in Figures~\ref{f:4p_delta2_fl} and
\ref{f:4p_delta2}.

The external field in the torus plane is given by
\begin{eqnarray}                                              
B_\mathrm{e}(R)&=& \frac{\mu_0}{2\pi}\left[
                   \frac{q_1L_1}{(R^2+L_1^2)^{3/2})}
                  +\frac{q_2L_2}{(R^2+L_2^2)^{3/2})}\right] \nonumber\\
               &=& \frac{\mu_0}{2\pi}\frac{q_1}{L_1^2}
                   \left[ \frac{1}{(\xi^2+1)^{3/2}}
                         +\frac{\epsilon\kappa^{-2}}
                               {(\xi^2\kappa^{-2}+1)^{3/2}}
                   \right],
\label{e:4p_Be}
\end{eqnarray}
where now $\xi=R/L_1$ and $\epsilon=q_2/q_1$, $\kappa=L_2/L_1$.
It has a decay index
\begin{equation}                                              
n_\mathrm{qp}=3\xi^2
 \frac{(\xi^2+1)^{-5/2}+\epsilon\kappa^{-4}(\xi^2\kappa^{-2}+1)^{-5/2}}
      {(\xi^2+1)^{-3/2}+\epsilon\kappa^{-2}(\xi^2\kappa^{-2}+1)^{-3/2}}\,.
\label{e:4p_n}
\end{equation}
This cannot be analytically solved for $\xi$ to obtain the threshold
radii corresponding to the critical decay index value (\ref{e:n_cr}).
The expressions for $R^2B_\mathrm{e}(R)$ and $\Psi_\mathrm{e}(R)$ are
fully analogous to (\ref{e:2p_R^2Be}) and (\ref{e:2p_Psie}), with
obvious extensions for the second pair of sources in the linear
quadrupole.

If the field strength or length-scale of the quadrupole are varied
with constant ratios $\epsilon$ and $\kappa$, then one expects the
system to behave in a similar manner with regard to the catastrophe as
in the case of the bipole field. This is verified below. Additionally,
we consider changes of $\epsilon$ or $\kappa$ leading to catastrophe.

\subsection{Changing the Source Strength}
\label{ss:4p_q}

\begin{figure}[t]                                             
\centering
\includegraphics[width=\columnwidth]{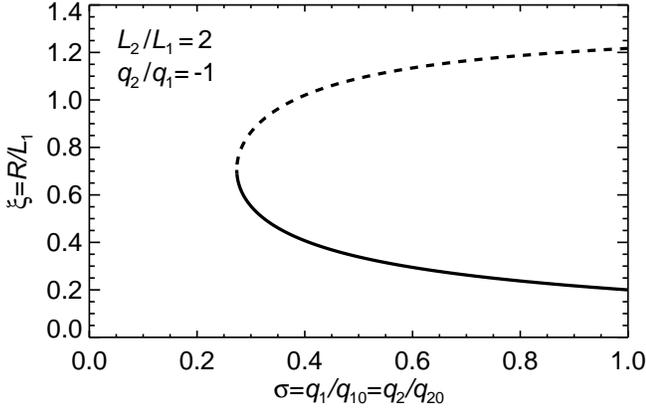} 
\caption{\label{f:4p_sigma}
 Equilibrium torus radius $R/L_1$ as a function of quadrupole strength
 $\sigma$ for an aspect ratio of $R/a=10$, size ratio $\kappa=2$, and
 charge ratio $\epsilon=-1$.}
\end{figure}

First we consider a proportional decrease of all four sources in the
linear quadrupole, $q_1(t)=\sigma(t)q_{10}$, $\epsilon=\mbox{const}$,
$L_1=L_{10}$, $\kappa=\mbox{const}$. Inserting the expressions for
$B_\mathrm{e}(R)$ and $\Psi_\mathrm{e}(R)$ into
Equation~(\ref{e:manifold}) again yields an explicit expression for
the equilibrium curve $\sigma=f(R,R_0/L_{1},R/a,\epsilon,\kappa)$,
\begin{equation}                                              
\sigma=
   \frac{2\xi_0^2[ (\xi_0^2+1)^{-3/2}
                  +\epsilon\kappa^{-2}(\xi_0^2\kappa^{-2}+1)^{-3/2}]}
        {\begin{array}{l}
          2\xi^2   [ (\xi^2+1)^{-3/2}
                    +\epsilon\kappa^{-2}(\xi^2\kappa^{-2}+1)^{-3/2}] \\
   \quad {}-(c_1/c_2)[ (\xi_0^2+1)^{-1/2}-(\xi^2+1)^{-1/2}           \\
              \quad {}+\epsilon(\xi_0^2\kappa^{-2}+1)^{-1/2}
                      -\epsilon(\xi^2\kappa^{-2}+1)^{-1/2}]
         \end{array}}\,,
\label{e:4p_sig_recon}
\end{equation}
but a closed analytical expression for the maximum of the denominator
can here no longer be obtained. The plot of this expression in
Figure~\ref{f:4p_sigma}, for the same value of the aspect ratio as in
Figure~\ref{f:2p_sigma} and for $\kappa=2$, $\epsilon=-1$,
demonstrates the expected fold catastrophe at
$(\sigma,R/L_{1})=(0.2738,0.7059)$, i.e., at a smaller radius than for
the external bipole field. At this radial position the field of the
linear quadrupole has a decay index of $n_\mathrm{qp}=1.413$, exactly
the threshold (\ref{e:n_cr}) of torus instability for the chosen
aspect ratio and $l_i$.

\begin{figure}[t]                                             
\centering
\includegraphics[width=\columnwidth]{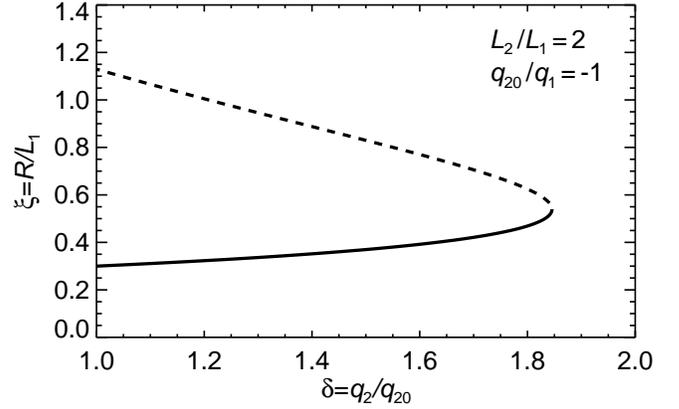} 
\caption{\label{f:4p_delta}
 Equilibrium torus radius $R/L_{1}$ as a function of relative strength
 of the source pairs in the linear quadrupole, measured by
 $\delta=\epsilon/\epsilon_0=q_2/q_{20}$, with the outer pair
 increasing in strength and otherwise the same parameters as in
 Figure~\ref{f:4p_sigma}.}
\end{figure}

Another path to catastrophe consists in varying the strength of only
one pair of sources in the quadrupole. We first let the relative
strength of the outer source pair increase as
$\epsilon(t)=\delta(t)\epsilon_0$ for opposite polarity
($\epsilon<0$), which decreases the external field in the torus plane
as well. The current channel is thus forced to find new equilibrium
positions at larger $R$ where $n_\mathrm{qp}$ is higher.
Equation~(\ref{e:manifold}) can again be easily solved for
$\delta=f(R,R_0/L_{1},R/a,\epsilon,\kappa)$. The resulting expression
\begin{equation}                                              
\delta=
 \frac{\begin{array}{l}
          2\xi_0^2 [ (\xi_0^2+1)^{-3/2}
                    +\epsilon_0\kappa^{-2}(\xi_0^2\kappa^{-2}+1)^{-3/2}] \\
     ~ {}-2\xi^2(\xi^2+1)^{-3/2}                                         \\
     ~ {}+(c_1/c_2)[(\xi_0^2+1)^{-1/2}-(\xi^2+1)^{-1/2}]
         \end{array}}
        {\begin{array}{l}
	  2\epsilon_0\xi^2\kappa^{-2}(\xi^2\kappa^{-2}+1)^{-3/2}]   \\
     ~ {}-\epsilon_0(c_1/c_2)[ (\xi_0^2\kappa^{-2}+1)^{-1/2}
                              -(\xi^2\kappa^{-2}+1)^{-1/2}]
	 \end{array}}
\label{e:4p_del_recon}
\end{equation}
is similar in structure to (\ref{e:4p_sig_recon}) and also requires a
numerical evaluation to demonstrate the catastrophe. Using $\kappa=2$
and $\epsilon=-1$ as in Figure~\ref{f:4p_sigma}, the catastrophe point
is found at $(\delta,R/L_{1})=(1.846,0.5396)$ where
$n_\mathrm{qp}=1.413$, again exactly at the threshold of the torus
instability (see Figure~\ref{f:4p_delta}). It lies in the radial range
of steeply increasing decay index below the magnetic null point at
$R/L_{1}=1.2$.

\begin{figure*}[t]                                            
\centering
\includegraphics[angle=-90,width=.7\linewidth]{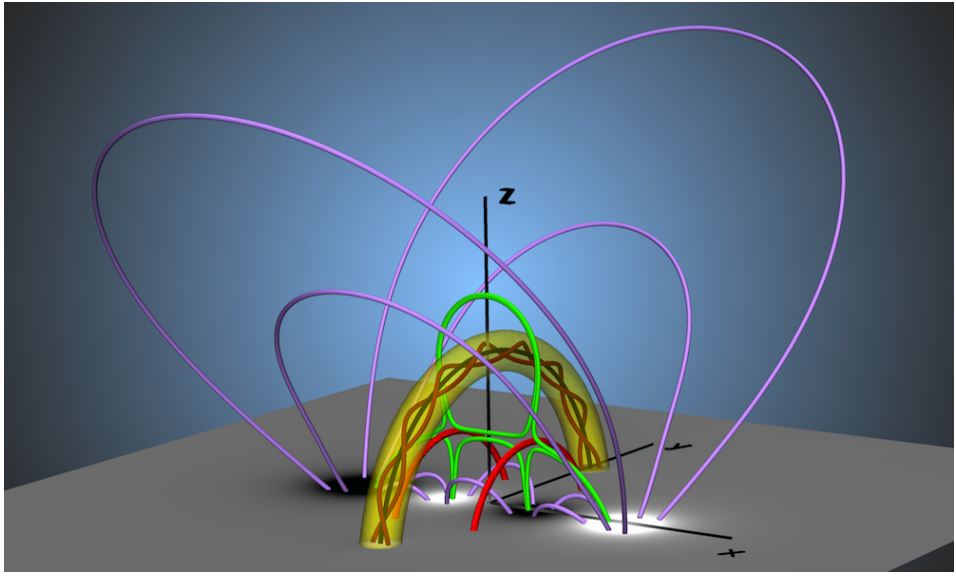}  
\caption{\label{f:4p_delta2_fl}
 Quadrupolar active-region model (generalized Titov-D\'emoulin
 equilibrium) shown in a format similar to Figure~\ref{f:2p_sigma_fl}
 for values of $R/a$ and $\kappa$ as in Figure~\ref{f:4p_delta2}.}
\end{figure*}

\begin{figure}[t]                                             
\centering
\includegraphics[width=\columnwidth]{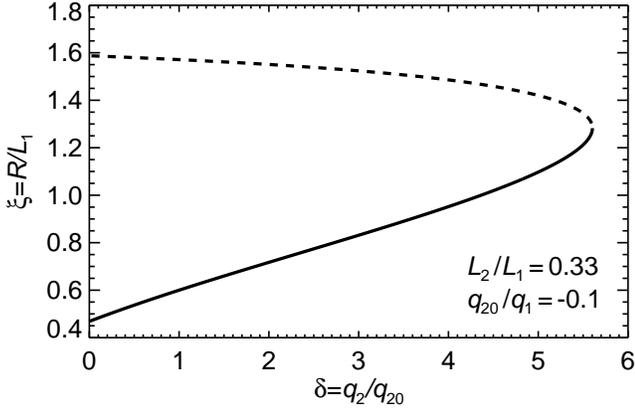} 
\caption{\label{f:4p_delta2}
 Equilibrium torus radius $R/L_{1}$ as a function of relative strength
 of the source pairs in the linear quadrupole, with the inner pair
 increasing in strength, size ratio $\kappa=1/3$, and aspect ratio
 $R/a=10$.}
\end{figure}

By placing the second bipole inside the first, $\kappa<1$, and
considering relatively small ratios of their source strengths,
$|\epsilon|=|q_2/q_1|<1$, we can address the influence of flux
emergence on the equilibrium, a process thought to be an efficient
trigger of eruptions \citep{Feynman&Martin1995}. Although the
dynamical behavior caused by reconnection between emerging and
preexisting flux is likely to play an important role in the
triggering \cite[e.g.,][]{Archontis&Hood2012, Kusano&al2012}, the
effects of the new flux on the force balance of the
current channel and on the decay index profile $n_\mathrm{qp}(R)$
alone can facilitate the transition to eruptive behavior.
Figures~\ref{f:4p_delta2_fl} and \ref{f:4p_delta2} show this for
flux emerging with an orientation anti-parallel to the main flux in
the region, $\epsilon_0=-0.1$, and $\kappa=1/3$. This
configuration contains two X-lines which do
not lie in the plane of the torus. Equations~(\ref{e:fc+source}),
(\ref{e:source}), (\ref{e:manifold}), and thus (\ref{e:4p_del_recon})
apply here as well, since reconnection will occur at the X-lines as
the torus expands, allowing it to ``slide through'' the external
poloidal field without changing the amount of enclosed flux.
Reconnected external flux is here transferred into the side lobes
under the X-lines instead of being added under the current channel.
The increase of the enclosed flux due to the emergence, which drives
the expansion of the torus, is described by the term
$\Delta\Psi_\mathrm{e}$ in Equation~(\ref{e:fc+source}). The torus
radius before flux emergence ($R/L_1\approx0.47$ for $\delta\to0$)
lies on the stable part of the equilibrium manifold (compare with
Figure~\ref{f:2p_sigma}). The emergence of anti-parallel flux weakens
the external poloidal field at the position of the current channel
(involving reconnection in the corona), so that the channel expands to
find a new equilibrium. Since now the profile $B_\mathrm{ep}(R)$ is
flatter in the range around the original position $R_0$, a catastrophe
occurs at a larger radius than in Figure~\ref{f:2p_sigma},
$(\delta,R/L_{1})=(5.603,1.283)$, but again exactly at the threshold
of torus instability: at this point $n_\mathrm{qp}=1.413$.

We did not find a catastrophe for $\kappa<1$ and $\epsilon$ increasing
from zero (modeling the emergence of flux with a parallel
orientation). An occurrence of catastrophe in this part of parameter
space requires the positive $\epsilon$ to decrease to a small value,
which weakens the external poloidal field in the plane of the torus,
as in all other cases considered in this paper. For completeness we
note that catastrophe and instability can be found for $\epsilon$
increasing from zero if the term $\Delta\Psi_\mathrm{e}$ is dropped in
expression~(\ref{e:fc+source}) for the enclosed flux. This changes the
relationship $I(R,p)$ and thus the balance between the hoop force
(quadratic in $I$) and the retracting force (linear in $I$) in
Equation~(\ref{e:Shafranov}), allowing the torus to expand in a range
of increasing small positive $\epsilon$ values. Since the new flux is
of smaller spatial scale, it raises the decay index and the expansion
leads to catastrophe, again at the threshold of torus instability.

\subsection{Changing the Length-scale}
\label{ss:4p_L}

\begin{figure}[t]                                            
\centering
\includegraphics[width=\columnwidth]{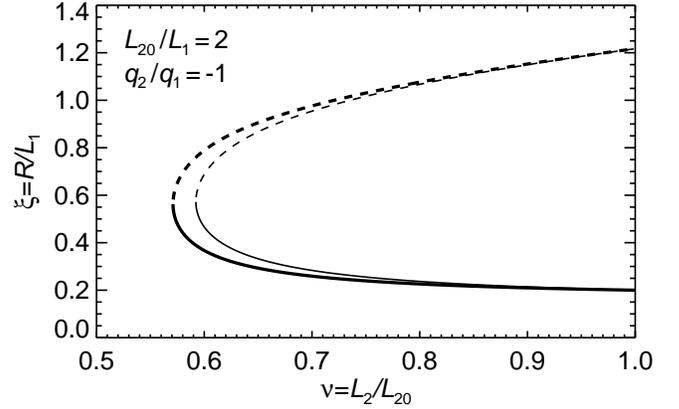} 
\caption{\label{f:4p_nu}
 Equilibrium torus radius $R/L_{1}$ as a function of the size ratio of
 the outer and inner source pairs in the linear quadrupole,
 measured by $\nu=\kappa/\kappa_0=L_2/L_{20}$, for the same parameters
 as in Figure~\ref{f:4p_sigma}. Excluding (including) the term
 $\Delta\Psi_\mathrm{e}$ in Equation~(\ref{e:manifold}) yields the
 equilibria on the thick (thin) line.}
\end{figure}

A proportional change of both length-scales in the linear quadrupole, 
$L_1(t)=\lambda(t)L_{10}$ with $\kappa=\mbox{const}$, has the same
effect as scaling the length-scale of the bipole field. The
equilibrium radius of the current channel changes proportionally to
$\lambda(t)$ if the term $\Delta\Psi_\mathrm{e}$ is dropped, and
neither instability nor catastrophe are reached in this case
(Equation~(\ref{e:manifold}), evaluated for this equilibrium, again
depends on $R/R_0$ and $\lambda$ only in the combination
$R/(\lambda R_0)$).

We thus consider the evolution driven by changing the size ratio of
the bipoles, $\kappa(t)=\nu(t)\kappa_0$, with all other parameters
held fixed, corresponding to a rearrangement of the flux distribution
in the photosphere. Similar to the increase of $|\epsilon|$ in
Figures~\ref{f:4p_delta} and \ref{f:4p_delta2}, an approach of $L_1$
and $L_2$ reduces the external field at the position of the current
channel. We show this for a decrease of $\kappa$ from the value used
in Figures~\ref{f:4p_sigma} and \ref{f:4p_delta}. The equilibrium
manifold (Equation~(\ref{e:manifold}) with the term
$\Delta\Psi_\mathrm{e}$ dropped) is given by
\begin{eqnarray}                                              
&& 2\xi^2   [ (\xi^2+1)^{-3/2}
     +\epsilon(\nu\kappa_0)^{-2}(\xi^2(\nu\kappa_0)^{-2}+1)^{-3/2}]\nonumber\\
&& {}-(c_1/c_2)[ 1+\epsilon-(\xi^2+1)^{-1/2}
     -\epsilon(\xi^2(\nu\kappa_0)^{-2}+1)^{-1/2}]          \nonumber\\
&&=  2\xi_0^2 [ (\xi_0^2+1)^{-3/2}
     +\epsilon\kappa_0^{-2}(\xi_0^2\kappa_0^{-2}+1)^{-3/2}]\nonumber\\
&&\phantom{=}
   {}-(c_1/c_2)[ 1+\epsilon-(\xi_0^2+1)^{-1/2}
     -\epsilon(\xi_0^2\kappa_0^{-2}+1)^{-1/2}]\,.          \nonumber\\
&&
\label{e:4p_lam_recon}
\end{eqnarray}
This implicit equation in both $\xi=R/L_{1}$ and $\nu$ must be
evaluated numerically. The result, plotted in Figure~\ref{f:4p_nu},
exhibits a fold catastrophe at $(\nu,R/L_{1})=(0.5710,0.5630)$ where
$n_\mathrm{qp}=1.413$, exactly at the threshold of torus instability.

\section{Summary and Conclusion}
\label{s:conclusion}

Using a toroidal flux rope embedded in a bipolar or quadrupolar
external field as a model for current-carrying coronal flux and its
associated image current, we have demonstrated the occurrence of fold
catastrophe by loss of equilibrium when magnetic reconnection can
proceed at an X-line under the flux rope. Several evolutionary
scenarios have been considered, which include changing the source
strength and length-scale of the external field. In each case, the
critical point for occurrence of the catastrophe coincides exactly
with the threshold for torus instability if the same or compatible
approximations are used, a result demonstrated to hold in general for
the adopted model. Catastrophe and torus instability are thus
equivalent descriptions for the onset of an eruption. They are based
on the same force balance for equilibrium and produce an onset of
eruption at the same point.

Thus, the merits of each description can be exploited while one can be
sure that the other description will yield the same onset point of
eruption. Analyzing an equilibrium for the occurrence of catastrophe
always includes a model for the pre-eruptive evolution and avoids the
consideration of unstable equilibria far away from the critical point,
which may be impossible to reach in reality. Analyzing the stability
of an equilibrium localizes the critical point without the need to
model the pre-eruptive evolution and in a formulation independent of
the specifics of such a model. Moreover, since only infinitesimally
small changes of the parameters must be considered in a stability
analysis, the adopted approximations may be better satisfied than
during the whole modeled pre-eruptive evolution in an analysis of
catastrophe. It is clear, however, that the approximations are equally
satisfied in the vicinity of the critical point.

\begin{acknowledgements}

We thank the referee for constructive comments which were very
helpful in improving this manuscript, and V.~S.~Titov for a careful
reading and commenting and for help in tracing the original work by
\citet{Osovets1959}.
B.K. acknowledges the hospitality of the solar group at Yunnan
Observatories, where part of his work was carried out, and the
associated support by the Chinese Academy of Sciences under grant
No.~2012T1J0017.
He also acknowledges support by the DFG, the STFC, and the NSF.
J.L.'s work was supported by 973 Program grants 2013CB815103 and
2011CB811403, NSFC grants 11273055, and 11333007, and CAS grant
KJCX2-EW-T07 to Yunnan Observatory.
E.R.P. is grateful to the Leverhulme Trust for financial support.
The contribution of T.T. was supported by NASA's HTP, LWS, and SR\&T
programs and by NSF.

\end{acknowledgements}


\bibliographystyle{apj}
\bibliography{ms_cat}

\end{document}